\documentclass[11pt,A4,fleqn]{article}
\usepackage[ruled,vlined,linesnumbered,noresetcount]{algorithm2e}
\usepackage{wrapfig,graphicx}
\usepackage{multirow}
\usepackage{amssymb}
\usepackage{amsthm,dsfont}
\usepackage[fleqn]{amsmath}
\usepackage{subfigure}
\usepackage{url,hyperref}
\usepackage[square,sort,comma,numbers]{natbib}
\usepackage{authblk}
\usepackage{anysize}
\usepackage[table,xcdraw]{xcolor}
\usepackage{float}
\restylefloat{table}
\makeatletter
\newcommand{\AlgoResetCount}{\renewcommand{\@ResetCounterIfNeeded}{\setcounter{AlgoLine}{0}}}
\makeatother

\marginsize{2cm}{2cm}{.3cm}{2cm}
\date{}
\begin{document}

\title{A supervised active learning method for identifying critical nodes in {W}ireless {S}ensor {N}etwork}
\author[1]{Behnam Ojaghi Kahjogh\thanks{bojaghi@uoc.edu}}
\author[2]{Mohammad Mahdi Dehshibi}

\affil[1]{Department of Computer Science, Universitat Oberta de Catalunya, Barcelona, Spain}
\affil[2]{Department of Computer Science and Engineering, Universidad Carlos III de Madrid, Legan\'{e}s, Spain}

\maketitle

\begin{abstract}
Energy Efficiency of a wireless sensor network (WSN) relies on its main characteristics, including hop-number, user's location, allocated power, and relay. Identifying nodes, which have more impact on these characteristics, is, however, subject to a substantial computational overhead and energy consumption. In this paper, we proposed an active learning approach to address the computational overhead of identifying critical nodes in a WSN. The proposed approach can overcome biasing in identifying non-critical nodes and needs much less effort in fine-tuning to adapt to the dynamic nature of WSN. 
This method benefits from the cooperation of clustering and classification modules to iteratively decrease the required number of data in a typical supervised learning scenario and to increase the accuracy in the presence of uninformative examples, i.e., non-critical nodes. Experiments show that the proposed method has more flexibility, compared to the state-of-the-art, to be employed in large scale WSN environments, the fifth-generation mobile networks (5G), and massively distributed IoT (i.e., sensor networks), where it can prolong the network lifetime.\newline

\end{abstract}

\section{Introduction}
Reasons such as gigantic data rate requirements, energy price, the ecological impact of carbon, and social responsibility for fighting climate change \cite{correia2010challenges} have inspired the telecommunication community to develop energy-saving techniques \cite{han2011green,earth3energy}. With the emergence of 5G technology, the importance of adopting energy-efficient architectures has been realized more to meet the demands of the increased capacity, improved data rate, and better quality of the service (QoS). Moreover, the primary concern of improving Energy Efficiency (EE) without compromising on user experience makes green communication an urgent need. Green communication not only can meet these demands but also can observe the social responsibility to reduce the carbon footprint by reducing the power consumption in a wireless network.

Taking an adequate EE metric into account can change the design and analysis of a power-optimized network. Energy Efficiency can be defined as either the number of bits transmitted per joule \cite{wu2015recent} or the system throughput for unit energy consumption \cite{gallager1987energy,verdu2002spectral}. Considering the application area, we can define Energy Efficiency as follows: (i) Information Theory: EE is minimizing the energy consumed per bit; (ii) Sensor Networks: EE is prolonging the network lifetime; and (iii) Cellular Networks: EE is prolonging the standby time of mobile station and maximizing EE under QoS constraint. In this paper, we focus on EE improvement in sensor networks. Bae and Stark~\cite{bae2009end} showed how hop-number, user’s location, and allocated power could influence on Energy Efficiency in a wireless sensor network. They mentioned that since the relay experiences fading conditions, it can be employed for energy saving. Using relay in energy saving schemes was introduced in the third generation of wireless communication. In 5G, this approach was extended to use relay along with coordination and support actions (CSA) and device-to-device (D2D) communication.

However, identify the best relay nodes is subject to a computational overhead which results in consuming much energy. Several studies tried to deal with computational overhead. Maric and Yates \cite{maric2010bandwidth} assumed that determining perfect global Channel state information (CSI) is feasible. However, this assumption is being too optimistic as imperfect CSI leads to degradation of performance in a wireless system \cite{gifford2005diversity}. Parzysz et al. \cite{parzysz2013energy} maximized the output signal-to-noise ratio for coherent and non-coherent relay networks considering imperfect CSI. The results reveal that, apart from individual and aggregate relay power constraints, ignoring uncertainties associated with global CSI often leads to poor performance. Selecting the relay node was the subject of \cite{zou2010adaptive,altubaishi2011variable} in which the least distance, highest transmission rate, and signal to noise ratio considered as selection schemes. Other parameters such as sensor coordinate, connectivity, and distance to the base station (BS) were not considered together with the relay in designing an energy-efficient WSN scheme to keep the computational overhead and, in turn, energy consumption as low as possible.

To address the computational overhead of identifying critical nodes in a wireless sensor network, we proposed to employ active learning method. Conventionally, a critical node defines as a node that its elimination will result in network disconnection \cite{dagdeviren2015semi}. However, in this study, we assumed that the network is strongly connected and can survive connectivity in the case of eliminating some critical nodes. In our definition, the removal of a critical node will increase network latency. We extended this definition to demonstrate how the proposed method can identify a critical node considering sensors’ coordinate, connectivity, distance to the base station (BS), and latency. Moreover, this assumption allowed us to compare the proposed active learning approach with other approaches that might fail or exceed computational limitations in finding an optimal solution. 

In a WSN scheme, nodes are configured in a dynamic environment, and the majority are non-critical. This dynamic and imbalance nature affects the design of the learning algorithm. To overcome the mentioned problems, we proposed an active learning algorithm to identify critical nodes. This method can overcome biasing in identifying non-critical nodes and needs much less effort in re-training to adapt to the dynamic nature. Compared to optimization-based solutions \cite{kahjogh2017impact,ojaghi2019sliced}, the proposed method has more flexibility to be utilized in large scale WSN environments. This active learning algorithm comprises two modules, i.e., clustering and classification modules. In the clustering module, we used local gravitation clustering (LGC) \cite{7915751} to reduce the imbalance impact. A linear support vector machine (SVM) \cite{cortes1995support} was used in the classification module to classify nodes into non-critical and critical classes. This active learning algorithm benefits from the cooperation of these modules to iteratively decrease the required number of data in a typical supervised learning scenario and to confront overwhelming by uninformative examples.

This paper is organized as follows. A review of the previous studies is given in Section~\ref{sec:survey}. In Section~\ref{sec:method}, we introduce the active learning algorithm, clustering, and classification modules. We explain in detail how the complexity of feature space can reduce the accuracy of the classification task. In Section~\ref{sec:experiment}, we examine analytics in the context of applications of interest. We conclude with concluding remarks and discussion of future work in Section~\ref{sec:conclusion}.

\section{Related Work} \label{sec:survey}
Various methods have been proposed to improve the EE in the present and future WSN schemes. Li \cite{li2015energy} contributed to relay selection and bandwidth exchange (BE). A heuristic energy-saving relay selection scheme was implemented to maintain minimum QoS of the user. Then, the surplus subcarriers were reallocated using a bandwidth exchange algorithm to save energy. The proposed algorithm with bandwidth exchange was compared with an algorithm without bandwidth exchange and a greedy algorithm. The results demonstrated that the proposed algorithm could reach maximum energy efficiency and minimum spectral efficiency. Parzysz et al. \cite{parzysz2013energy} proposed a relay power allocation method to maximize the output SNR, observing individual and aggregate relay power constraints. Results revealed that ignoring uncertainties of CSI can significantly reduce network performance.

In \cite{ashraf2011sleep}, they developed an optimal power control method in which a stochastic approach was utilized to establish a trade-off between EE and fairness. In this algorithm, power cost and channel conditions played the primary roles in deciding the priority of each relay. A relay node can then be selected if all nodes with higher priority were already selected. They employed a routine similar to the least mean square algorithm to control the power. Results showed that using a convex optimization problem considering the fairness criterion for each relay can maximize network lifetime. Zhang et al. \cite{zhang2010power} used a game-theoretic approach to allocate powers among all the active nodes in order to maximize the network sum-rate of relays and source EE. In this approach, a reinforcement learning technique was used to calculate the Q-values of different agents. A feasibility check was also performed over the user’s QoS constraint. In the case of infeasibility, the admission control was passed on to the BS observing the QoS requirement of nodes. Results showed that the full-duplex mode outperforms the half-duplex in terms of both energy efficiency and network sum-rate. The system’s EE increased by raising the number of relays or falling the number of sources.

Chang et al. \cite{chang2016energy} proposed an algorithm consisting of (1) a fast selection scheme for the active antenna subsets at the relay, and (2) a user selection and transmission power optimization process to maximize the energy efficiency in a multiple-input and multiple-output relay system. Experimental results showed that the algorithm could achieve optimal EE and high-gain even in low SNR regimes with low complexity. Wang et al. \cite{wang2014cellular} proposed an energy-efficient maximum weighted matching algorithm to improve the EE of a system with multi-relay, multi-user orthogonal frequency division multiple access. In this algorithm, Dinkelbach’s method was used to solve the optimization problem by solving a sequence of subtractive concave problems. Optimization objective was met by matching source users to relay users in order to increase the capacity and to improve the performance of source and cell-edge users, respectively. From the experimental results, they concluded that when there is insufficient power for attaining maximum EE, the algorithm reaches the upper bound and does not make use of additional available power. They also showed that energy efficiency decreased by the increase of the SNR threshold.

Several research groups and associations have been studied the impact of other network’s features, besides relay nodes, on energy efficiency. Mobile VCE has focused on the BS hardware, architecture, and operation, resulting in energy savings of 75–92 percent in simulations \cite{mobilevce}. EARTH has developed an array of new technologies, including low-loss antennas, micro direct transmission, antenna muting, and adaptive sectorization, according to traffic fluctuations, realizing energy saving gains of 60–70 percent with less than 5 percent throughput degradation \cite{skillermark2012enhancing}. Several operators have been actively developing and deploying green technologies, including green BSs powered solely by renewable energies, and green access infrastructure such as cloud/collaborative/clean radio access network (C-RAN) \cite{mobile2011c}. 

C-RAN implemented a soft and virtualized BS with multiple baseband units integrated as virtual machines on the same server, supporting multiple radio access technologies. A soft end-to-end solution from the core network to the RAN can enable the 5G goals of spectral and energy efficiency. Kahjogh et al. \cite{yildiz2017impact,icufn} showed that the elimination of some nodes (i.e., critical nodes) has a significant impact on lifetime and latency. They proposed Sequential and Bulk algorithms to identify critical nodes. In the sequential algorithm, the impact of removing each node from the network was analyzed, and in the bulk algorithm, critical nodes were identified by solving an optimization problem. The major drawback of these methods was the computational complexity, resulting in identifying at most two critical nodes. Following their idea, we put one step forward and proposed an active learning approach to resolve this drawback, having the capability of identifying the majority of critical nodes and the flexibility of adapting to the dynamic nature of the WSN with reasonable time complexity.

\section{Proposed Method} \label{sec:method}
In a wireless communication network, removing a critical node can result in network disconnection and steep rising in latency. In this study, we assumed that the network is strongly connected and can survive connectivity in the case of eliminating some critical nodes. We defined a critical node as a node that its removal increases network latency. Considering the dynamic nature of wireless communication, finding critical nodes is not an offline problem, and the computational overhead of using optimization methods \cite{sepas2012novel,dehshibi2017hybrid} can substantially decrease network lifetime. Although substituting classification algorithms with optimization methods can confront computational complexity, the imbalance problem (where only 15\% of nodes are critical ones) can result in a high false-positive rate. 

In this study, we proposed to employ an active learning approach to identify critical nodes. To resolve the imbalance problem, we utilized local gravity clustering (LGC) method \cite{7915751}, where the cluster with the maximum size is then selected as a new problem space. In this way, the new problem space contains nodes which 30\% of them are critical ones. Compared to the original problem space, we observed a 50\% increase in critical nodes’ cardinality and a 15\% decrease in non-critical's. The advantage of LGC over clustering approaches like k-means is that it is not dependent on the initial seed and can observe the problem characteristics. We classified nodes by using a linear support vector machine (SVM) \cite{cortes1995support} with the Leave-One-Out cross-validation configuration to identify critical nodes. The proposed policy that links the clustering and classification modules could decrease the computational complexity of re-training, bring flexibility, and increase the generalizability. Details of the proposed method are as follows.

\RestyleAlgo{ruled}
\SetAlgoVlined
\LinesNumbered
\AlgoResetCount
\begin{algorithm}[!htbp]
\SetKwInOut{Input}{Input}\SetKwInOut{Output}{Output}
\Input{$\mathcal{L}$: Labeled set,\\
        $\mathcal{U}$: Unlabeled set, \\ 
        $m$: Training set size.}
\Output{Model $\mathcal{M}$}
	\While{training size $\leq m$}{
		$\mathcal{M} \leftarrow$ learn a model based on $\mathcal{L}$ \\
		$\mathcal{U} \leftarrow \mathcal{X} \setminus \mathcal{L}$ \\
		\ForEach{$q_{i} \in \mathcal{U}$}{	
			$u_{i} \leftarrow Loss(q_{i})$
			}
		$u^{*} \leftarrow \arg\min_{i}(u_{i})$ \\
		$\mathcal{L} \leftarrow \mathcal{L} \cup u^{*}$ \\
		$\mathcal{U} \leftarrow \mathcal{U} \setminus u^{*}$ \\
		$\mathcal{U} \leftarrow $ update the model based on $\mathcal{L}$
	}
\Return{$\mathcal{M}$}
\caption{Active Learning Algorithm.}
\label{alg:M1}
\end{algorithm}

We divided our problem space ($\mathcal{X}$) into two subsets referred to as (1) labeled ($\mathcal{L}$) and (2) unlabeled ($\mathcal{U}$) sets. A node is represented by a 5-tuple $(c_x, c_y, Con, Dist, Rel)$ with a corresponding label $\mathcal{Y} = \{0,1\}$, where `1' is critical and `0' otherwise. In this representation, $(c_x, c_y)$ refer to the node's coordinate and $Con, Dist$, and $Rel$ are connectivity, distance to the BS, and relay, respectively. The learning model $\mathcal{M}$ maps the input to output, and a loss function $Loss(\mathcal{M})$ measures the model’s error. At each learning iteration, the learner selects a query $q \in \mathcal{U}\subset \mathcal{X}$ and obtains the label $l \in \mathcal{Y}$. Adding $q$ to the previously labeled instances $\mathcal{L}$, and re-training the model $\mathcal{M}$ with this new pair $(q, l)$ results in reducing loss value. We used the entropy regularization instead of minimizing the expected error to satisfy the self-learning condition. In this way, we could also use the latent structure to improve $\mathcal{M}$ (see Algorithm \ref{alg:M1}).

We utilized LGC to (1) create $\mathcal{L}$ and $\mathcal{U}$ sets, and (2) score $l$. We selected two clusters with the largest number of samples. It was observed in the course of experiments that the first cluster ($\mathcal{L}$) comprises 80-83\% of critical nodes, and the second cluster ($\mathcal{U}$) contains 15-17\% of the critical nodes, with the ratio of 30:70 for critical to non-critical nodes in each set. If the union of both sets does not contain all critical nodes, we count this difference in calculating the final error. In the LGC, physical laws of gravitation were used to define the relationship between data points and their neighborhoods. In this algorithm, the local resultant forces are first calculated for each data point. Data points are then classified as interior, boundary, and unlabeled points. Afterward, the SoftConnection routine is applied to those points which were labeled as interior points. Finally, the data points labeled as boundary points are assigned to the clusters. This procedure is described in Algorithm \ref{alg:M2}, where details of the SoftConnection routine can be found in \cite{wang2018clustering}.

\RestyleAlgo{ruled}
\SetAlgoVlined
\LinesNumbered
\AlgoResetCount
\begin{algorithm}[!htbp]
    \SetKwInOut{Input}{Input} \SetKwInOut{Output}{Procedure}
    \Input{$\mathcal{X}$: Data set,\\
            $k$: Number of neighbors,\\ 
            $\widehat{CE}$:Centrality threshold,\\
            $\mathbf{IM}$: Initial momentum}
    
    \Output{$\mathcal{L}$: Labeled set}	
    \ForEach{$x_i \in \mathcal{X}$}{	
    	$m_{i} = \frac{1}{\sum_{j=1}^{k}D_{ij}}$ \\
    	$\overrightarrow{F_{i}} = \frac{1}{m_{i}}\sum_{j=1}^{k}\hat{D}_{ij}$
    }
    \ForEach{$x_i \in \mathcal{X}$}{	
    	$CE_i = \frac{\sum_{j=1}^{k}\cos \left (\overrightarrow{F}_j,\overrightarrow{D}_{ji} \right )}{k}$ \\
    	$CO_i = \sum_{j=1}^{k}\left (  \overrightarrow{F}_i\cdot\overrightarrow{F}_i \right )$
    }	
    SoftConnection($\mathcal{X}, \widehat{CE}, \mathbf{IM}$) \\
    \ForEach{un-clustered boundary point $x_i$}{	
    	\If {$CO_i \geq 0$}{
    	    Add $x_i$ to its nearest cluster it points at.
    	 }
    }	
    \ForEach{un-clustered boundary point $x_i$}{	
    Add $x_i$ according to its neighbors' cluster labels.
    }	
    \Return{$\mathcal{L}$}
\caption{LGC Algorithm. \cite{wang2018clustering}}
\label{alg:M2}
\end{algorithm}

In addition to real labels, the hidden cluster’s label $k \in \{1, 2, \cdots, K\}$ is also introduced where $K$ is the number of clusters. $k$ indicates that the sample belongs to the $k^{th}$ cluster, where all information about the class label $y$ is already encoded in $k$. Therefore, by knowing $k$, $x \in \mathcal{X}$ and $y \in \mathcal{Y}$ can be considered independent, and the joint distribution is written as Eq. \ref{eq:M1}:
\begin{equation}
    \label{eq:M1}
    p(x,y,k) = p(y|x,k)p(x|k)p(k) = p(y|k)p(x|k)p(k)
\end{equation}

In our model, $p(y|k)$ is defined as a logistic regression for all clusters with the same parameters (Eq. \ref{eq:M2}).
\begin{equation}
    \label{eq:M2}
    p(y|k) = \frac{1}{1+e^{-y(c_{k}.a+b)}}
\end{equation}
where $c_{k}$ is the representative of the $k^{th}$ cluster, $a \in \mathbb{R}^{d}$ and $b \in R$ are the logistic regression parameters. Therefore, once all the parameters of $p(y|k)$ are determined, this likelihood can be used to determine the label of samples belong to this cluster. Given $p(y|k)$, the posterior probability of sample label can be calculated by Eq. \ref{eq:M3}
\begin{equation}
    \label{eq:M3}
    p(y|x) = \sum_{k=1}^{K}p(y,k|x) = \sum_{k=1}^{K}p(y|k)p(k|x)
\end{equation}
where $ p(k|x)= p(x|k)p(k)/p(x)$. The new label for data is subject to clustering and can be defined by using the Bayes decision rule (\ref{eq:M4}):
\begin{equation}
\label{eq:M4}
\hat{y}(x) = \begin{cases} 1 & \text{if} \;  p(y=1|x;\hat{a},\hat{b}) > p(y=0|x;\hat{a},\hat{b}) \\ 0 & \text{otherwise} \end{cases}
\end{equation}
where $\hat{a},\hat{b}$ denote the current estimates of the parameters. Equation \ref{eq:M3} shows that the weighted combination of the posterior probability is used for the representatives. Moreover, the same label will be assigned to well-clustered instances as the nearest representative.

\begin{figure}[!htbp]
    \centering
    \subfigure[]{\includegraphics[width=0.6\textwidth]{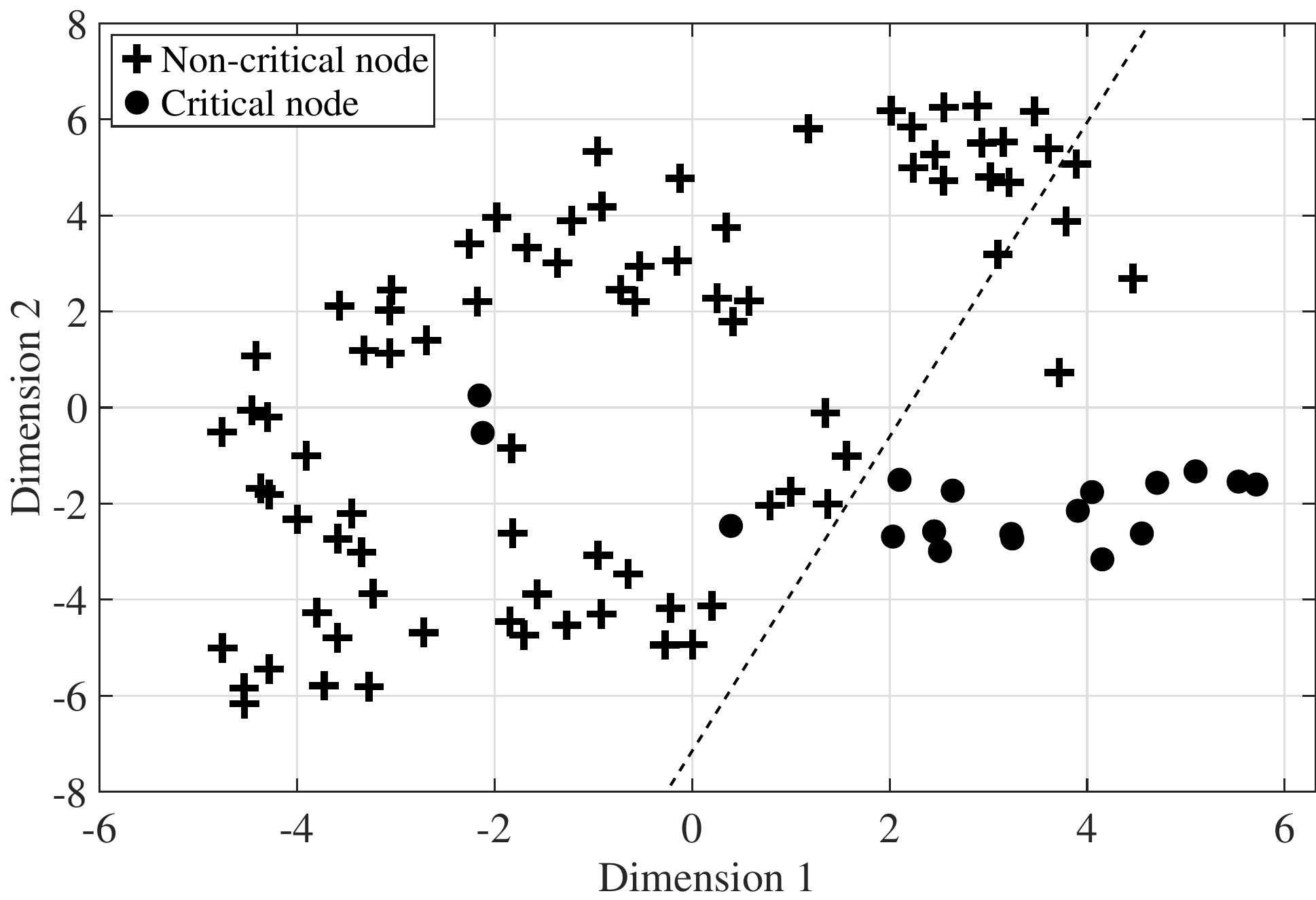}}
    \subfigure[]{\includegraphics[width=0.6\textwidth]{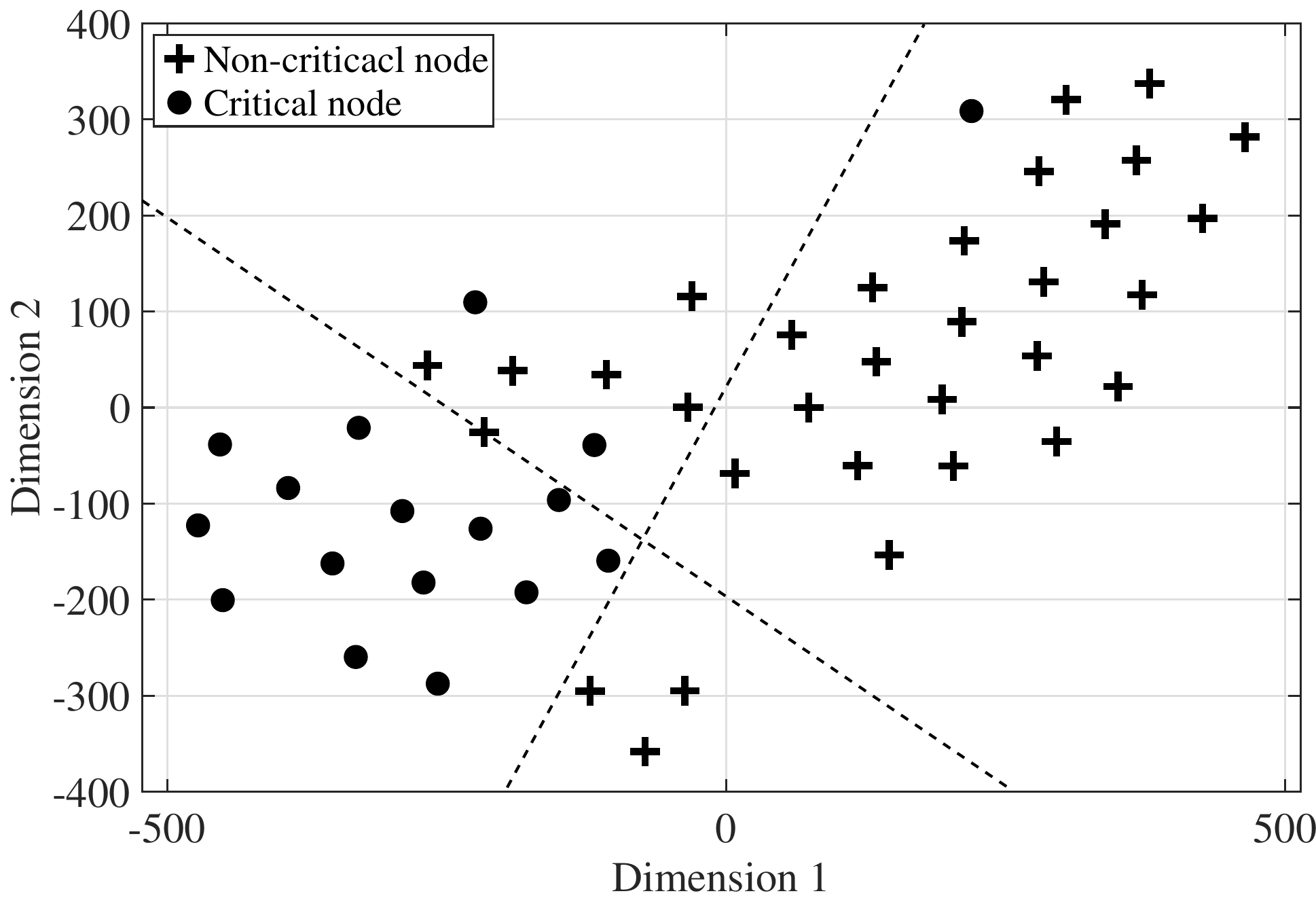}}
    \caption{Distribution of latent space by applying t-SNE (a) original network configuration, (b) selected nodes by applying LGC.}
    \label{fig:M1}
\end{figure}

We visualized the feature space using t-distributed stochastic neighbor embedding (t-SNE) \cite{maaten2008visualizing} to show the impact of clustering. t-SNE is a nonlinear dimensionality reduction technique that is used for visualizing high-dimensional latent space in two or three dimensions. In this approach, similar objects are modeled by nearby points and distant points with high probability model dissimilar objects. Figure \ref{fig:M1}(a) shows the distribution of the original feature space considering all nodes. Figure \ref{fig:M1}(b), however, shows the distribution of feature space where the most informative clusters are considered. 

In Figure \ref{fig:M1}, we synthetically added dashed lines to show how a classifier can discriminate feature space using support vectors. It is apparent that classifying $\mathcal{X}$ will result in a high false-positive rate due to the biasing problem that is resolved after clustering feature space.

Following the clustering process, we used a support vector machine (SVM) \cite{cortes1995support} with a linear kernel to identify critical nodes. This classification algorithm tries to minimize the objective function of Eq. \ref{eq:M5} to find the maximum margin hyperplane parameters dividing the two classes.

\begin{align}
    \label{eq:M5}
    \min_{\overrightarrow{w},b,\xi,C} \frac{1}{2}\left \| \overrightarrow{w} \right \|^{2}+C \sum_{i=1}^{l}\xi_{i} \nonumber \\
    \text{subject to:} \nonumber \\
    \forall_{i=1}^{l}:y_{i}[\overrightarrow{w}.\Phi (\overrightarrow{q_{i}})+b] \geq 1-\xi_{i} \nonumber \\
    \forall_{i=1}^{l}: \xi_{i}>0
\end{align}
where $\overrightarrow{w}$ and $b$ are the parameters defining the maximum margin hyper-plane, $\Phi (\cdot)$ is the kernel function, $C$ is the parameter defining the trade-off between the margin size and mis-classified examples, $q_i$ is a sample from $\mathcal{L}$, $y_i$ is the associated label to $q_i$, and $\xi$ is the slack variable.

\section{Experimental Results} \label{sec:experiment}
To prove the efficiency of the proposed method in identifying critical nodes, we deployed a WSN with 100 nodes and ran experiments over this scheme. In the course of experiments, clustering and classification modules of the proposed active learning approach were compared to state-of-the-art ones. We showed that the proposed method outperforms optimization approach in identifying critical nodes in terms of accuracy and computational complexity. We also examined the importance of nodes' characteristics in the performance of the proposed method.

\subsection{Simulated Scenario}
\begin{wrapfigure}{r}{0.5\textwidth}
    \centering
    \includegraphics[width=0.4\textwidth]{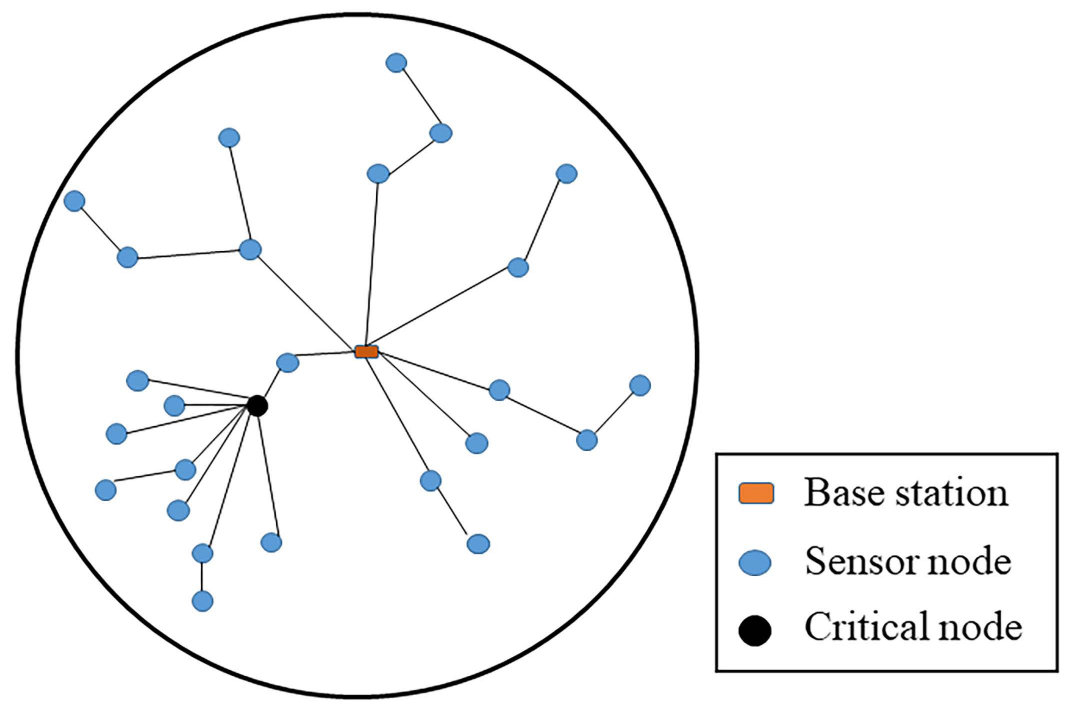}
    \caption{A schematic view of the network.}
    \label{fig:B1}
\end{wrapfigure}
Without loss of generality, we deployed a disk-shaped network topology with a radius of 200~m, including $N_{S}=100$ nodes, where a single base station placed at the center and $N_{S}-1 = 99$ sensors distributed within the disk. We used a uniform random distribution to determine the positions of sensor nodes which have a maximum transmission range of $R_{max} = 82.92$~m as defined in~\cite{ashraf2011sleep}. In this topology, schematically shown in Figure~\ref{fig:B1}, each sensor covers a target (i.e., neighbor node) if the Euclidean distance between the sensor and the target is less or equal than $R_{max}$. We used GAMS\footnote{General Algebraic Modeling System (GAMS). Available: \url{http://www.gams.com/}.} to construct a sensor network. Then, we passed `$(c_x,c_y)$-coordinates,' `connectivity,' and `distance to base station' to the mixed integer programming (MIP) framework to find `relay' and `latency' using CPLEX solver. All computations have been run on an Intel Core i7 4702 MQ, 2.20 GHz up to 3.20 GHz processor with 8 GB DDR3 memory.

\subsection{An optimization approach to identify critical nodes}
Besides the active learning approach, we also proposed to employ a mixed-integer programming (MIP) approach to identify critical nodes with two goals: (1) providing ground-truth for data fed into the proposed active learning approach; (2) comparing the proposed learning-based approach with an optimization approach in terms of computational complexity and flexibility to adapt to the network changes.

In this approach, a MIP model is used to evaluate the criticality of each network’s node. For the given topology, the MIP runs for $N_{S}$ times, and the node that its removal results in the maximum latency adds to critical nodes set. In our implementation, we used the summation of flows in the network as a metric for calculating the total hop count. In Algorithm~\ref{alg:B1}, the network topology is represented by $G=(V, E)$, where $V$ is the node, and $G$ is the edge. This algorithm takes the number of critical nodes to be found ($N_{C}$) as an input and returns an ordered set $(C,\leq_{latency})$ of critical nodes associated with the latency. This algorithm iterates for $N_{C}$ times, and in each iteration finds a node that its removal from $G$ maximizes the network latency. Considering the total iteration of  $\begin{pmatrix} N_{S} \\ N_{C} \end{pmatrix}$, the time complexity of this algorithm is $O(N_{C}N_{S})$

\RestyleAlgo{ruled}
\SetAlgoVlined
\LinesNumbered
\AlgoResetCount
\begin{algorithm}[!htbp]
    \SetKwInOut{Input}{Input}\SetKwInOut{Output}{Output}
    \Input{$G$=($V$, $E$): Network topology graph \\
    	  $N_{C}$: Number of critical nodes}
    \Output{$(C,\leq_{latency}) = \{(v_{j}, lt_{j}) | v_j \in V, \; 1 \leq j \leq n, \; lt_j: \text{Network latency by removing } (v_{BS}, v_j) \text{ from } G\}$}
    \For{k=1 to $N_{C}$}{
    	$C_k.latency=0$ \\
    	$C_k.latency=\infty$
    }
    \ForEach{$v_i~\epsilon~V, \; 1\le~i\le|V|$}{
    	$lt_i \leftarrow~ latency(G\backslash v_i)$ \\
    	\If {($lt_i > C_k.latency$)}{
    		$C_k.latency = lt_i$
    	}
    }
    $G \leftarrow G \backslash C_k.criticalNode$\\
    \tcc{Nodes that their removal give the maximum latency are removed from $G$}
    \Return{C}.
\caption{Critical Node Identification}
\label{alg:B1}
\end{algorithm}

\subsection{Performance Evaluation}
\begin{figure}[!htbp]
    \centering
    \includegraphics[width=0.4\textwidth]{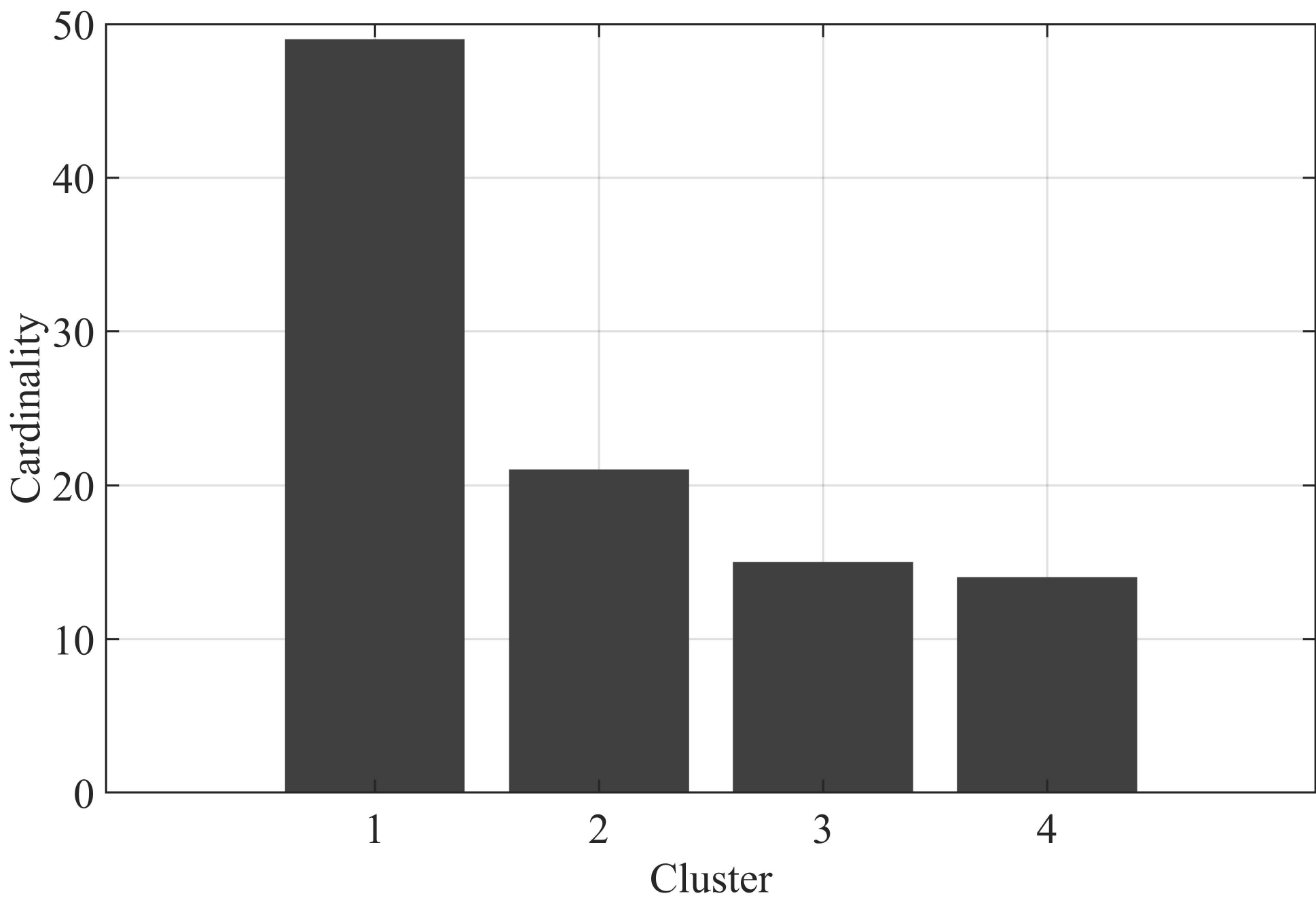}
    \caption{Results of applying LGC to the deployed network. Four clusters are formed, where Cluster 1 contains $N_{C}-1$ critical nodes, and Cluster 2 contains another critical node.}
    \label{fig:M3}
\end{figure}

In Figure~\ref{fig:M1}, we showed that utilizing LGS can reduce feature space complexity. In the deployed network, the local gravity clustering method partitioned the problem spaces into four clusters, as is shown in Figure~\ref{fig:M3}.

We observed that Cluster 1 contains $N_{C}-2=18$ critical nodes, and Cluster 2 has the remaining critical nodes. Therefore, these two clusters are selected as the labeled ($\mathcal{L}$) and unlabeled ($\mathcal{U}$) set. The cardinality of the new problem space is reduced in a way that 30\% of nodes are critical ones. Compared to the original problem space, we observed a 1.33 times increase in critical nodes’ cardinality and a 15\% decrease in non-critical ones, which helps in solving the imbalance problem. It is worth mentioning that there is a probability of appearing one critical node in other clusters. However, the error of overlooking this instance is much less than the error of biasing in the classifier as a result of increasing the number of non-critical nodes in the training set. To show the efficiency of utilizing LGC, we also clustered problem space with the k-means \cite{aghaahmadi2013clustering}, where the initial seed was set to 2, 3, and 4, respectively. Table \ref{tbl:clust} shows the distribution of nodes in each cluster. It is obvious that applying k-means brings subjective results, which makes it inappropriate to be used in the proposed active learning approach.

\begin{table}[H]
\centering
\caption{Distribution of critical nodes in different clusters as the result of applying the k-means clustering approach.}
\label{tbl:clust}
\begin{tabular}{lllll}
\hline
                  & \textbf{Cluster 1} & \textbf{Cluster 2} & \textbf{Cluster 3} & \textbf{Cluster 4} \\ \hline
\textbf{seed = 2} & 12/76              & 8/34               & NA                 & NA                 \\
\textbf{seed = 3} & 5/21               & 9/54               & 6/25               & NA                 \\
\textbf{seed = 4} & 0/30               & 6/25               & 2/15               & 12/30              \\ \hline
\end{tabular}
\end{table}

The classification module of the proposed active learning approach is equipped with a linear SVM~\cite{suykens1999least}. To demonstrate the efficiency of the proposed active learning approach in identifying critical nodes, we compared it with the case that a linear SVM solely applied to the total problem space $\mathcal{X}$. Table~\ref{tbl:class} tabulates confusion matrices for both proposed and compete methods.

\begin{table}[H]
\centering
\caption{Results of utilizing the proposed active learning approach and a linear SVM in identifying critical nodes.}
\label{tbl:class}
\begin{tabular}{lllllll}
 & \multicolumn{2}{l}{\multirow{2}{*}{\textbf{}}} & \multicolumn{2}{c}{\textbf{Predicted class}} & \multirow{2}{*}{\textbf{TP rate}} & \multirow{2}{*}{\textbf{FN rate}} \\ \cline{4-5}
 & \multicolumn{2}{l}{} & \textbf{Critical} & \textbf{Non-critical} &  &  \\ \hline
\multirow{2}{*}{\textbf{Active learning}} & \multirow{4}{*}{\textbf{True class}} & \multicolumn{1}{l|}{\textbf{Critical}} & 15 & 4 & 78.94\% & 21.06\% \\
 &  & \multicolumn{1}{l|}{\textbf{Non-critical}} & 1 & 69 & 98.58\% & 1.42\% \\
\multirow{2}{*}{\textbf{Linear SVM}} &  & \multicolumn{1}{l|}{\textbf{Critical}} & 8 & 12 & 40.00\% & 60.00\% \\
 &  & \multicolumn{1}{l|}{\textbf{Non-critical}} & 2 & 78 & 97.5\% & 2.5\% \\ \hline
\end{tabular}
\end{table}

Due to the dependency of clustering and classification modules in the proposed method and imbalance nature of the classification task, we must define a weighted accuracy (Eq.~\ref{eq:per1}) and report extra measures for a fair comparison. In addition to true positive and false negative rates, we calculated weighted precision, recall, and F1-score (Eq.~\ref{eq:mf1}). A weighted precision computes the precision independently for each class and then take the weighted average. Weighted recall is the fraction of the total amount of relevant instances that were actually retrieved weighted by the weight of each class. $F1_{weighted}$ score is a harmonic mean of precision and recall.
\begin{equation}
\label{eq:per1}
\text{weighted accuracy} = \sum_{k=1}^{|\mathcal{Y}|} w_i \sum_{x: y = k} I\left(y = \hat{y}\right)\,.
\end{equation}
where $I$ is the indicator function and returns 1 if the classes match and 0 otherwise. $w_k$ is the assigned weight to each class such that $\sum_{k=1}^{|\mathcal{Y}|} w_k = 1$. The higher the value of $w_k$ for an individual class, the greater is the influence of observations from that class on the weighted accuracy. In this study, we set $w_{0}=0.7, w_{1}=0.3$. 

\begin{gather} \label{eq:mf1}
    P = \frac{TP}{TP+FP}, \; R = \frac{TP}{TP+FN}, \; F1 = 2P\frac{R}{(P + R)}, \nonumber\\
    P_{weighted} = \sum_{k=1}^{|\mathcal{Y}|} w_i P_i, \; R_{weighted} = \sum_{k=1}^{|\mathcal{Y}|} w_i R_i, \; F1_{weighted} = \sum_{k=1}^{|\mathcal{Y}|} w_i F1_i.
\end{gather}
where $TP, FP, TN$, and $FN$ stand for true positive, false positive, true negative, and false negative, respectively. Table~\ref{tbl:score} shows the comparison of both methods in terms of evaluation measures.

\begin{table}[!htbp]
\centering
\caption{The comparison of Proposed method and Linear SVM in terms of evaluation measures}
\label{tbl:score}
    \begin{tabular}{l|lll}
    \hline
     & \textbf{$P_{weighted}$} & \textbf{$R_{weighted}$} & \textbf{$F1_{weighted}$} \\ \hline
    \textbf{Proposed method} & 84.83\% & 61.11\% & 91.70\% \\
    \textbf{Linear SVM} & 57.25\% & 41.63\% & 69.14\% \\ \hline
    \end{tabular}
\end{table}

\begin{figure}[!htbp]
    \centering
    \includegraphics[width=0.55\textwidth]{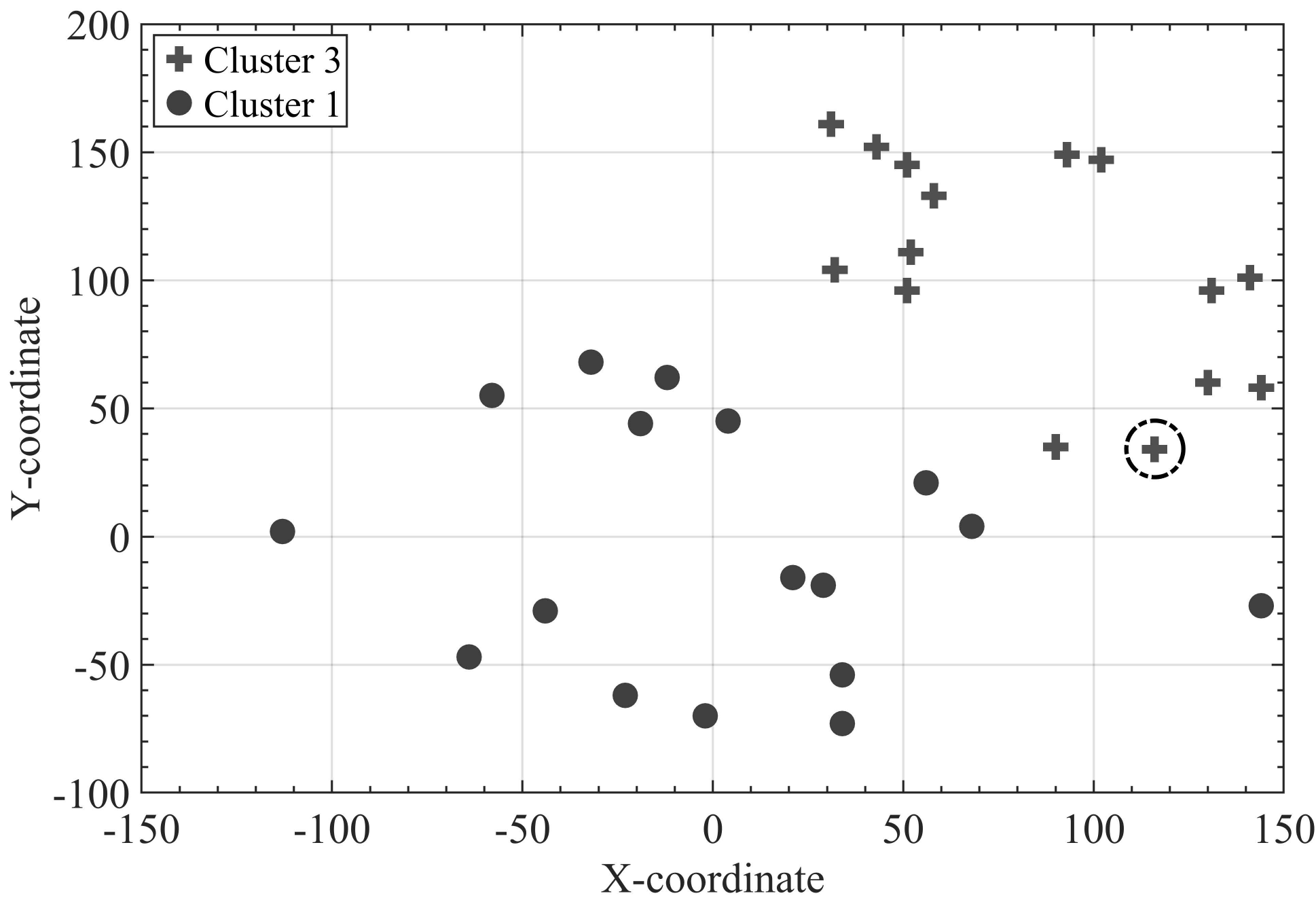}
    \caption{Scatter plot of nodes' location. $\bullet$ and $+$ are used to point to Cluster 1 and Cluster 3 instances, respectively. The dashed circle highlights the node that is a critical node and was assigned to Cluster 3.}
    \label{fig:M2}
\end{figure}

In our experiments, we observed that LGC assigned one critical node to Cluster 3, resulting in a new type of error. We plotted nodes according to their $(c_x,c_y)$-coordinates to see if the coordinate can affect this assignment. Figure~\ref{fig:M2} shows the scatter plot of nodes that are clustered in Cluster 1 and Cluster 3. We selected these two clusters as Cluster 1 contains the majority of critical nodes, and the majority of instances in Cluster 3 are non-critical nodes. It is evident that the the $(c_x,c_y)$-coordinate cannot solely change the distribution of clusters. Otherwise, there are points in both clusters that have an approximately equal Euclidean distance and could be misassigned.

\begin{table}[H]
\centering
\caption{Cluster 3 instances, where the highlighted row shows the critical node.}
\label{tbl:3}
\begin{tabular}{cccccc}
\hline
 & \multicolumn{2}{c}{\textbf{Coordinate}} &  &  &  \\ \cline{2-3}
\multirow{-2}{*}{\textbf{ID}} & \textbf{x} & \textbf{y} & \multirow{-2}{*}{\textbf{Connectivity}} & \multirow{-2}{*}{\textbf{Dist-to-BS}} & \multirow{-2}{*}{\textbf{Relay}} \\ \hline
6 & 144 & 58 & 6 & 154.7094 & 1 \\
18 & 32 & 104 & 16 & 109.0566 & 2 \\
19 & 90 & 35 & 9 & 96.87996 & 3 \\
21 & 43 & 152 & 12 & 157.748 & 1 \\
22 & 31 & 161 & 13 & 164.1266 & 1 \\
61 & 102 & 147 & 10 & 178.7874 & 1 \\
71 & 131 & 96 & 10 & 162.1654 & 1 \\
76 & 141 & 101 & 6 & 172.7776 & 1 \\
78 & 52 & 111 & 13 & 123.1841 & 3 \\
79 & 93 & 149 & 10 & 175.0799 & 1 \\
85 & 51 & 96 & 15 & 108.8205 & 2 \\
90 & 51 & 145 & 10 & 154.032 & 1 \\
91 & 58 & 133 & 11 & 145.1135 & 1 \\
93 & 130 & 60 & 6 & 143.6442 & 1 \\
\rowcolor[HTML]{F56B00} 
95 & 116 & 34 & 10 & 120.5214 & 4 \\ \hline
\end{tabular}
\end{table}

We compared features of this node with the nodes assigned to Cluster 1. Tables~\ref{tbl:3} and~\ref{tbl:4} show characteristics of instances belong to Cluster 1 and Cluster 3, respectively. Highlighted row in Table~\ref{tbl:3} shows the critical node assigned to Cluster 3 and highlighted rows in Table~\ref{tbl:4}show nodes that could be candidates for assigning to Cluster 3.

\begin{table}[H]
\centering
\caption{Instances belong to Cluster 1 where the highlighted rows are the possible candidates to be assigned to Cluster 3.}
\label{tbl:4}
\begin{tabular}{cccccc}
\hline
 & \multicolumn{2}{c}{\textbf{Coordinate}} &  &  &  \\ \cline{2-3}
\multirow{-2}{*}{\textbf{ID}} & \textbf{x} & \textbf{y} & \multirow{-2}{*}{\textbf{Connectivity}} & \multirow{-2}{*}{\textbf{Dist-to-BS}} & \multirow{-2}{*}{\textbf{Relay}} \\ \hline
7 & -19 & 44 & 15 & 48.24708 & 5 \\
13 & -64 & -47 & 12 & 79.78816 & 7 \\
26 & 34 & -73 & 14 & 80.31338 & 11 \\
28 & -58 & 55 & 16 & 80.23978 & 7 \\
31 & -23 & -62 & 15 & 66.10476 & 9 \\
36 & -44 & -29 & 15 & 52.63024 & 13 \\
38 & -12 & 62 & 14 & 63.59438 & 5 \\
39 & 56 & 21 & 11 & 59.38667 & 8 \\
\rowcolor[HTML]{96FFFB} 
59 & 68 & 4 & 9 & 68.59481 & 6 \\
\rowcolor[HTML]{9AFF99} 
64 & 144 & -27 & 11 & 146.8928 & 5 \\
68 & -32 & 68 & 15 & 75.38416 & 14 \\
70 & 34 & -54 & 14 & 63.92478 & 4 \\
75 & 21 & -16 & 12 & 26.56586 & 3 \\
80 & -113 & 2 & 13 & 113.0047 & 2 \\
88 & -2 & -70 & 13 & 70.26103 & 3 \\
99 & 4 & 45 & 14 & 45.28009 & 3 \\
100 & 29 & -19 & 12 & 34.84477 & 1 \\ \hline
\end{tabular}
\end{table}

In our deployment, the maximum transmission range is set to 82~m to relay data to the neighbor nodes. Distance to the base station for nodes 59 and 95 are, therefore, in the range that transmission can be done with two hops, whereas node 59 needs one hop to transmit data. The connectivity of these three nodes is almost similar. In terms of the relay, however, node 95 has a relay value less than nodes 64 and 59. LGC algorithm, first, considers the distance to the base station and selects nodes with higher values as non-critical nodes candidates. Then, it checks relay and connectivity features and selects nodes with lower values as a probable candidate for clusters with the majority of non-critical nodes. As a result, the features of node 95 are in a way that this error type is unavoidable. However, it is apparent from the evaluation criteria that the proposed method not only can cover this error but also can handle the imbalance nature of data adequately.

\section{Conclusion} \label{sec:conclusion}
With the emergence of 5G technology, the importance of adopting energy-efficient architectures has been realized more to meet the demands of the increased capacity, improved data rate, and better quality of the service (QoS). Considering the application area, Energy Efficiency defined in three categories, which are Information Theory, Sensor Networks, Cellular Networks. In this paper, we focused on energy efficiency in sensor networks to prolong the network lifetime. 

Different WSN characteristics such as hop-number, user’s location, allocated power, and relay can influence Energy Efficiency in a wireless sensor network, which identifying them is subject to a substantial computational overhead and consuming much energy. To address the computational overhead of identifying critical nodes in a wireless sensor network, we proposed to employ an active learning approach. In this study, we assumed that the network is strongly connected and can survive connectivity in the case of eliminating some critical nodes. In our definition, the removal of a critical node will increase network latency. The proposed active learning approach can overcome biasing in identifying non-critical nodes and needs much less effort in retraining to adapt to the dynamic nature of WSN. 

The proposed method benefits from the cooperation of clustering and classification modules to iteratively decrease the required number of data in a typical supervised learning scenario and to confront overwhelming by uninformative examples. Compared to the state-of-the-art methods, experimental results show that the proposed method has more flexibility to be utilized in large scale WSN environments.

\bibliographystyle{plain}
\bibliography{manuscript.bib}

\end{document}